\documentclass{sig-iptps}
\usepackage{epsfig}
\usepackage{times}
\usepackage{cite}
\usepackage{color}
\usepackage{graphicx}
\usepackage{subfigure}
\usepackage{fancyhdr}
\usepackage{url}

\long\def\com#1{}
\begin{document}

\title{\vspace{-0.9em}User-Relative Names for\\ Globally Connected Personal 
       Devices\vspace{-0.5em} }
\author{Bryan Ford, Jacob Strauss, Chris Lesniewski-Laas, \\
	Sean Rhea, Frans Kaashoek, and Robert Morris \\
	Massachusetts Institute of Technology}
\date{}

\maketitle

\thispagestyle{fancy}

\begin{abstract}
Nontechnical users
who own increasingly ubiquitous network-enabled personal devices
such as laptops, digital cameras, and smart phones
need a simple, intuitive, and secure way
to share information and services
between their devices.
{\em User Information Architecture}, or UIA,
is a novel naming and peer-to-peer connectivity architecture
addressing this need.
Users assign UIA names
by ``introducing'' devices to each other on a common local-area network,
but these names remain securely bound to their target as devices migrate.
Multiple devices owned by the same user,
once introduced,
automatically merge their namespaces
to form a distributed {\em personal cluster}
that the owner can access or modify from any of his devices.
Instead of requiring users to allocate globally unique names
from a central authority,
UIA enables users to assign their own {\em user-relative} names
both to their own devices and to other users.
With UIA, for example,
Alice can always access her iPod from any of her own personal devices
at any location
via the name {\tt ipod},
and her friend Bob can access her iPod
via a relative name like {\tt ipod.Alice}.
\end{abstract}

%
% $Id: intro.tex,v 1.17 2006/02/12 19:42:44 jastr Exp $
%

\section{Introduction}

\vspace{1.5ex}

Personal devices such as 
mobile phones, 
digital music players,
personal digital assistants,
console gaming systems, and digital cameras
are now ubiquitous in the lives of ordinary people.
As these devices proliferate,
peer-to-peer connectivity between them
is increasingly important.
For example,
a user may copy photos from a camera to a PC for storage,
to a web page for publishing,
or to a photo iPod to take on the road,
and perhaps from there to a friend's iPod.
One current transfer mechanism---plugging 
devices together via USB cable---is both
straightforward and secure:
the cable itself physically indicates
which devices should participate in the transfer,
and the isolated physical medium guarantees its security.

As personal devices begin to support wireless networking
and Internet connectivity,
we would like to extend the simplicity and security
of a USB cable to device connectivity on a global scale.
Alice should be able to connect her WiFi-enabled iPod
to her home PC via a ``virtual cable,''
so that she can 
browse photos or play music stored there
from a WiFi-enabled coffee shop or friend's house.
Setting up this ``virtual cable'' should not require
technical knowledge or special configuration on Alice's part,
and it should continue working
even when the devices it connects are
behind firewalls or NATs.

If Alice meets Bob in a coffee shop,
she should easily be able to share with him
information or services located on any of her personal devices.
Bob should be able to connect to Alice's devices
even after he leaves the coffee shop,
until she chooses to sever their relationship.
No one else
should be able to impersonate Bob,
however,
in order to gain access to Alice's shared resources.

%To make this global peer-to-peer sharing model work,
%users need need a simple, intuitive, and secure way
%to name their own personal devices and those of their friends.
%Grandma should be able to build and use her network of personal devices
%without knowledge of technical details
%such as IP addresses, dynamic DNS, or public and private keys.
The {\em User Information Architecture}, or UIA,
is a peer-to-peer connectivity architecture
that provides users a simple, intuitive, and secure way
to share information and services
between personal devices
by assigning ad hoc names that act like ``virtual cables.''
Users assign names
by ``introducing'' devices to each other on a common network.
%such as a home network or a coffee shop's WiFi network.
Unlike the ephemeral names used in rendezvous schemes
such as Apple Bonjour~\cite{bonjour}, however,
UIA names persist and remain securely bound to
the global cryptographic identities
of their targets~\cite{moskowitz03hip-arch,mazieres99separating,rivest96sdsi}
as devices migrate.
Once Alice introduces her iPod to her home PC,
her iPod can continue accessing her PC
by the same name
from anywhere she finds Internet access.

%UIA does not require globally unique names.
In a network of billions of users,
globally unique names would inevitably have to
look something like \url{ipod.alicesm5186.myisp.com},
substantially limiting their conciseness and readability.
%of being short, human-readable, and easy to remember.
UIA names are instead {\em user-relative}:
users control their own private namespaces
much as they control their mobile phones' address books today.
Unlike a conventional address book, however,
a UIA namespace is shared across
all the devices a user owns:
changes made on one device
automatically propagate to the others.

Users assign user-relative UIA names not only to their own devices
but also to other users.
Bob might know Alice as ``Alice'',
her company directory might list her as ``Alice Smith, Marketing'',
and her son might simply name her ``Mom''.
If Alice gives Bob access to some files on her PC,
he accesses them via a name
analogous to ``Alice's PC''.
In this way, UIA adapts peer-to-peer social networking ideas
previously explored for other purposes~\cite{turtle,sprout,sybildht}
to form a secure peer-to-peer naming infrastructure.

\com{

We believe our results have applicability beyond the domain of personal
devices to other peer-to-peer systems.  For example, systems such as
SFR~\cite{sfr} and $i3$~\cite{i3} also identify network endpoints using
cryptographically secure, but otherwise meaningless, bit strings.  Our
system can be used to provide human-readable names to such endpoints in
a secure and user-friendly manner.

}%com

\com{
Several peer-to-peer systems have recently leveraged pre-existing social
relationships to improve the security of peer-to-peer
routing~\cite{turtle,sprout,sybildht}.  UIA presents a new use of these
relationships, using them to build secure naming as well.
}

The next section presents the goals of UIA's naming system, and 
Section~\ref{sec-exp} describes its operation % the operation of UIA
from a non-technical user's viewpoint.
Section~\ref{sec-design} develops the technical details of UIA's design,
and Section~\ref{sec-impl} summarizes implementation status.
Section~\ref{sec-related} presents related work,
and Section~\ref{sec-concl} concludes.

\section{Goals of UIA}

The purpose of UIA
is to provide users with a convenient and intuitive way
to name and communicate with their own and their friends' personal
devices.
%to name their and their friends' personal devices,
%in order to enable both local and remote access to them
%by themselves and their trusted acquaintances.
%To succeed in this purpose,
To this end, UIA must satisfy the following goals:
\begin{itemize}
\setlength{\itemsep}{0em}
\setlength{\topsep}{0em}

\item	Names must be {\em user-relative}
	%that are directly meaningful to themselves,
	so as not to require global uniqueness.
	If Alice owns only one laptop and has only one friend named Bob,
	she should be able to refer to them % on her devices
	as {\tt laptop} and {\tt bob}, % respectively,
	despite the % fact that the world contains
	millions of other laptops and people named Bob in the world.

\item	Names must have {\em strong bindings}
	to the identities of the objects named,
	independent of their current physical location
	or network attachment point.
	When Alice refers to her name {\tt laptop},
	the name should always resolve
	to {\em her} laptop or fail to resolve % at all
	(e.g., if it is turned off); %  or not on the network),
	no other device should be able to impersonate it.

\item	Assigning names
	must be a {\em simple and intuitive} process.
	If Alice meets Bob at a conference
	and their laptops share a common WiFi network,
	assigning a name to Bob should be as simple as
	looking him up in a local network browser window
	and clicking ``Bookmark''.

\item	A user % who owns many personal devices
	should only have to manage {\em one namespace}.
	If Alice already owns several devices,
        % a laptop, a desktop PC, and a camera,
	% then purchases a new iPod,
	she should only have to name a newly purchased device once,
	not once on each existing device.

\item	Users should easily be able to {\em share}
	their personal namespaces and device resources selectively
	with trusted friends and acquaintances.
	If Alice gives Bob permission to access
	some files on her desktop PC, % back at home,
	he should have access to them
	via a name as simple as ``Alice's PC''.

\item	When physically possible,
        UIA should {\em automatically provide connectivity} among devices
        that have naming relationships, including 
        %When the user refers to a personal name,
	%UIA should automatically {\em find a way to communicate}
	%with the named target whenever physically possible.
	%Communication should function smoothly
	%across the Internet, 
        between Internet-connected private LANs
	and within ad hoc networks disconnected from the Internet
	(e.g., among passengers in an airplane).

\item	Finally, UIA 
	%personal names 
        should {\em coexist cleanly} with DNS,
	% with the Internet's existing global namespace,
	so that a user can %freely 
        use personal names like {\tt laptop}
	alongside global names like {\tt amazon.com}
	within the same application.
\end{itemize}

\section{User Experience}
\label{sec-exp}

\begin{figure}[t]
\centering
\includegraphics[width=0.47\textwidth]{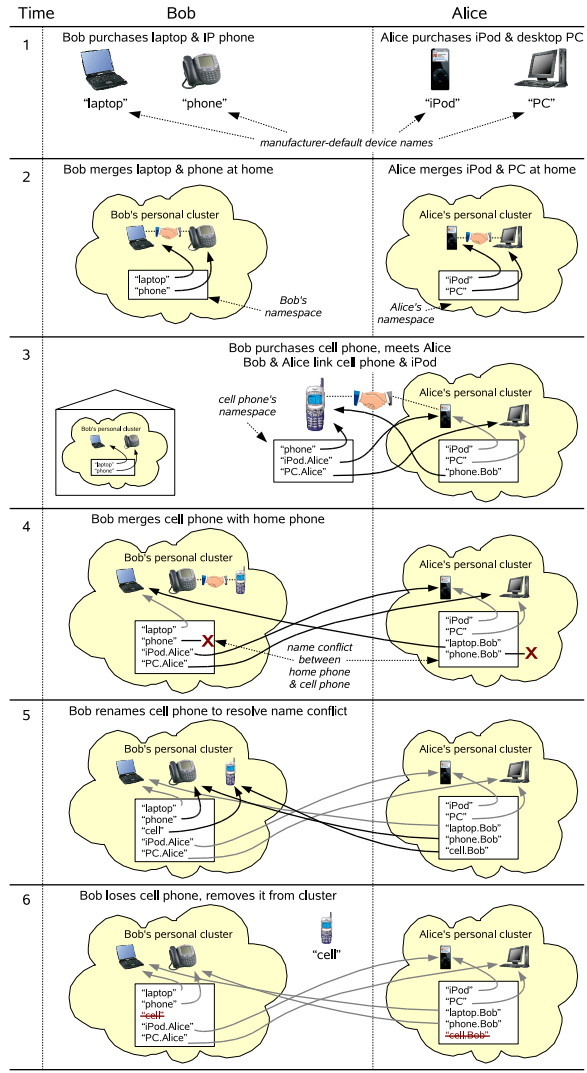}
\caption{Example Personal Device Scenario}
\label{fig:story}
\end{figure}

This section describes UIA's key operating principles
from the perspective of a non-technical user,
demonstrating how it satisfies the goals listed above
with the help of an example scenario illustrated in Figure~\ref{fig:story}.
Technical details of how the system provides this user experience
follow in the next section.

\subsection{Local Introduction, Remote Access}

% NOTE: Frans found it confusing that we overload the term
% "introduce" to refer both to merging and linking
% namespaces.  I'm trying to clarify it.  Here is the
% original text:  --srhea
%
% UIA devices normally acquire names for each other
% when users {\em introduce} them to each other
% on a local-area network.
% These names then persist on the devices
% and subsequently facilitate {\em remote access}
% as they migrate to different networks.
% Users can introduce UIA devices in two different ways:
% one way is intended for multiple devices owned by the same user,
% the other for devices owned by different users.
% 
% Each device ideally ships from its manufacturer
% UIA-enabled and pre-configured with a generic name for itself
% that the user can accept or change as desired.
% A user who owns several devices
% can introduce them so as to merge their logical namespaces
% and form a distributed {\em personal device cluster}.

Each UIA device ideally ships from its manufacturer
UIA-enabled and pre-configured with a generic name for
itself such as {\tt laptop} or {\tt phone}
that the user can accept or change as desired.  After
purchase, the device acquires names for other devices as its
user \emph{introduces} it to them on a local-area network.  These
introductions assign persistent names by which one device
can securely refer to the other, and these names
subsequently facilitate remote access as the devices migrate
to different networks.  

Users can introduce UIA devices in two different ways: they
can \emph{merge} two or more of their own devices to form a
\emph{personal device cluster} sharing a common logical namespace,
and they can create named
\emph{social links} from their own clusters to other users' personal users.

\subsection{Personal Device Clusters}

At Time~1 in the scenario shown in Figure~\ref{fig:story},  
Bob purchases and brings home a new laptop and Internet phone,
having default names {\tt laptop} and {\tt phone} respectively.
At Time~2 Bob uses UIA's local rendezvous tool on each device,
similar to those already available with Bonjour~\cite{bonjour},
to find the other device on his home WiFi network
and selects the UIA ``Merge Devices'' command on each.  This
action merges the two devices into a single personal device
cluster, allowing names already known to
or subsequently entered on each device to be used
on the other device as well.

While working on his laptop, for example,
Bob can now refer to his phone by its
manufacturer-configured name, {\tt phone}.
Furthermore, the merge operation is persistent and secure;
when Bob subsequently takes his laptop on a trip, he can
remotely access his home phone from his laptop anywhere he
finds Internet access, to check his voice messages for
example, still using the same name, {\tt phone}.

\subsection{Name Sharing and Social Networking}

Users can also create named links from their personal
device clusters to those of other users
in order to reflect social relationships
and facilitate information sharing.
With this second form of introduction,
users retain exclusive control
over their clusters' private namespaces,
but can selectively grant access to known acquaintances,
without making their devices vulnerable to unknown hosts
that may inhabit the same physical network.

In the example scenario,
Bob purchases a new WiFi-enabled cell phone at Time~3
and meets Alice at a coffee shop on his way home,
before he has merged his cell phone
with his other devices.
Bob finds Alice's iPod using his cell phone's local rendezvous tool
and selects UIA's ``Link to Another User's Device'' command,
and Alice does likewise.
Bob's phone presents Alice's self-chosen user name if available,
such as {\tt Alice},
as a suggestion for Bob's new name for Alice,
but Bob can override this hint as desired---%
using an alternative like {\tt Alice Smith} or {\tt Alice from IPTPS}
for example
if Bob already knows other Alices.
If Alice now grants Bob access to some files on her desktop PC at home
(which is named {\tt PC} in Alice's personal namespace),
then Bob can subsequently access
these files from his cell phone
under the ``Bob-relative'' name {\tt PC.Alice}.

Bob then returns home and merges his new cell phone
with his home phone,
as shown at Time~4 in Figure~\ref{fig:story}.
Bob's laptop transitively learns
about the cell phone's membership in the cluster
without him having to merge them explicitly,
and Alice can now name Bob's laptop on her devices
via the Alice-relative name {\tt laptop.Bob}.

\subsection{Resolving Conflicts}

Unfortunately, both of Bob's phones
happened to have identical default names of {\tt phone},
resulting in their names conflicting
in his newly merged namespace.
UIA notifies Bob of the conflict,
and he can continue using the non-conflicting name {\tt laptop}
but must resolve the conflict before the name {\tt phone} will work again.
Bob resolves the conflict
on his cell phone at Time~5,
by renaming it {\tt cell}
while leaving the home phone with the name {\tt phone}.
Bob's other devices learn the resolved name bindings automatically,
and Alice's devices now see Bob's phones
as {\tt phone.Bob} and {\tt cell.Bob}.

If Bob makes conflicting namespace changes
on two of his devices
while they are disconnected from each other,
UIA similarly detects the conflict once the devices reconnect.
Bob can continue using other names
while conflicts exist,
and he can resolve such conflicts at leisure on any of his devices.

\subsection{Removing Devices}

When Bob loses his cell phone at Time~6,
he uses his laptop to remove the cell phone from his personal cluster.
The laptop notifies Bob's other devices
as well as Alice's,
protecting them from subsequent access via the cell phone.
If Bob's phone was stolen
and Bob does not remove it from his cluster promptly,
the thief could gain remote access
to Bob's other devices,
illustrating one inherent risk of greater connectivity.

% Frans thinks this sentence is unnecessary:  --srhea
%Login mechanisms or biometric security features on portable devices
%can help mitigate this risk.

\section{Design}
\label{sec-design}

% 
% $Id: record-types.tex,v 1.6 2006/02/13 03:19:37 srhea Exp $ 
% 

\begin{table*}
\begin{center}
\small
\begin{tabular}{|c|c|l|}
\hline
Record Type & Fields & Purpose \\
\hline
\hline
\emph{create namespace} &
 &
creates a new empty namespace 

\\ \hline

\emph{link} &
\emph{parent pointer},

& 
maps \emph{name} to the namespace \emph{child pointer} in the
namespace pointed to 

\\
& \emph{child pointer}, \emph{name} &
by \emph{parent pointer}
\\ \hline

\emph{name device} &
\emph{parent pointer}, &
maps \emph{name} to the device \emph{device EID} in the namespace pointed to \\
& \emph{device EID}, \emph{name} 
& by \emph{parent pointer} \\
\hline
\emph{merge} &
\emph{local pointer}, &
merges the namespaces pointed to by \emph{local pointer} and
\emph{remote pointer} \\
& \emph{remote pointer} &
so that they share device names and social links placed into either one\\
\hline
\emph{unlink} &
\emph{pointer} &
unmaps the namespace linked in the record 
pointed to by \emph{pointer} \\
\hline
\emph{remove name} &
\emph{pointer} &
removes the device name created in the record 
pointed to by \emph{pointer} \\
\hline
\emph{stop merge} &
\emph{pointer}, \emph{stop seq.}, &
stops importing new records from the namespace imported at \emph{pointer} \\
&\emph{stop hash} 
& after the record with sequence \emph{stop seq.} and hash \emph{stop
hash} \\
\hline
\end{tabular}
\end{center}
\caption{\label{tbl-log-records}\emph{The primary types of log records in UIA.}
In addition to the fields shown, each
record contains the device EID, a sequence number, and the secure hash
of the record that preceded it.  Pointers to records are implemented as
the triple of these three fields.  
}
\label{table:types}
\end{table*}

This section outlines the elements of UIA's design that are key to realizing
the user experience described above.

\subsection{Device Identity and State Management}

To give UIA devices strong, decentralized identities,
each device
hashes the public key of a locally-generated key pair to construct its
\emph{endpoint identifier}, or EID.
As in similar identity
schemes~\cite{rivest96sdsi,mazieres99separating,moskowitz03hip-arch},
EIDs are
cryptographically unique, self-configuring, and self-certifying,
but not human-readable.

To manage user-friendly names and their associations with secure EIDs,
a UIA device generates {\em change records}
in response to user actions affecting the namespace
and stores these records in an append-only per-device log.
Each device signs the change records it creates,
and devices with namespace sharing relationships replicate each other's
logs via gossip.  
To resolve user-relative names, a device examines both the records in
its own log and those in its replicas of other device's logs; no network
communication is performed during name resolution.
Replicating logs in this manner guards users' state against
device loss or failure and keeps the namespace available during periods
of disconnection.

Table~\ref{table:types} summarizes the essential information contained in
the most common types of change records.
This table is not intended to be complete or definitive,
but merely to illustrate the general design strategy.
Additional information and record types may be needed in the future,
for example,
to support more sophisticated namespace sharing and access control policies.

%\note{I don't like that parenthetical.  We should be more direct and
%just say we plan to extend it but aren't sure about the details yet.
%--srhea}

\subsection{Name Resolution}

To resolve UIA names, each device maintains a tree representing the names
embodied in its log and in its replicas of other devices' logs.
The internal nodes
of this tree consist of \emph{create namespace} records that have been
transitively merged via \emph{merge} records, and the
branches of the tree consist of named \emph{link} records that join
these merged namespaces.
Each device designates a
single \emph{create namespace} record as the root of its local tree.

For compatibility with DNS,
UIA names follow the same formatting rules as DNS names,
consisting of a series of {\em labels} separated by dots.
To resolve a specific name such as {\tt PC.Alice} on Bob's phone,
for example,
Bob's phone first parses the name into its component labels,
{\tt PC} and {\tt Alice}.
Starting from its own root namespace,
Bob's phone then traverses its namespace tree
by following successive links corresponding to the labels in the name,
working from right to left as in DNS:
from Bob's root namespace to Alice's root namespace
via the label {\tt Alice},
then from Alice's root namespace to Alice's PC
via the label {\tt PC}.

The remainder of this section clarifies the above rough summary
of the name resolution process,
by following the sequence of events
shown in Figure~\ref{fig:story}.

\subsection{Naming Devices and Building Clusters}

To implement the preconfigured name for Bob's VoIP phone in
Figure~\ref{fig:story}, UIA writes a \emph{create namespace} record 
at the beginning of the phone's log, designates it as the root of the
phone's namespace, then writes a \emph{name device} record with the
phone's EID, a pointer to the root namespace record, and the name
{\tt phone}.  In this way, the phone's local name for itself is just
{\tt phone}: the name {\tt phone} is contained in the \emph{name device}
record, and there are no named \emph{link} records between it and the
root.

When users introduce devices to form clusters,
each device writes a \emph{merge} record
merging its local root with that of the other.
\emph{Both} devices must write, digitally sign,
and exchange corresponding \emph{merge} records
in order either device to consider the merge process complete.
If a device encounters an unpaired \emph{merge} record,
which could happen due to a hardware failure or loss of connectivity
during a merge, for example,
the device treats the unpaired \emph{merge} record as a ``conflict''
that the user can resolve at leisure by completing or canceling the merge.

When Bob introduces his laptop to his phone at Time~2 in the figure,
for example,
the two devices merge each other's root namespaces.  Since their root
namespaces are now linked by corresponding \emph{merge} records, they are
treated as the same node in the logical namespace tree,
and each device can now refer to
the other via \emph{name device} records
taken from either original namespace.  The laptop, for
example, refers to the phone simply as {\tt phone}, since that is the name
in the phone's \emph{name device} record, and there are no named
\emph{link} records between it and the newly-merged root. 

\subsection{Introducing and Naming Users}

UIA uses the \emph{link} record type to
provide access from one user's devices to those of another.  When Alice
and Bob link their devices at Time~3 in the figure, Alice's iPod
links Bob's root namespace into her own by writing a \emph{link} record
with the name {\tt Bob} and pointers to the two relevant namespaces.
Bob's phone similarly mirrors these actions,
linking its own root namespace to the
iPod's root namespace via the name {\tt Alice}.

Alice can
now refer to Bob's phone, for example,
as {\tt phone.Bob}.
To resolve this name,
Alice's iPod follows its new \emph{link} record for the label {\tt Bob},
from Alice's root namespace to Bob's root namespace,
then from Bob's namespace to Bob's phone
via Bob's original \emph{name device} record for {\tt phone},
originally written on Bob's phone
but gossiped to Alice's iPod after the introduction.

\subsection{Merging Clusters}

At Time~4 in the figure,
Bob returns home and merges his new cell phone with his home
phone.
Bob's logical root namespace,
and in effect Bob's ``user identity'' for UIA's purposes,
is now transitively defined
according the set of valid {\em merge} record pairs
that link together the root namespaces of Bob's devices:
namely the {\em merge} pair Bob generated earlier at time~2,
and the new {\em merge} pair from Time~4.
Bob's laptop and cell phone thus
discover each other through gossip with the home phone and begin
gossiping together in turn.  Likewise, Bob's laptop and home phone
learn of Alice's devices through his cell phone, and Alice's devices
similarly learn of Bob's additional devices.

Although Alice's existing \emph{link} record for {\tt Bob}
only directly contains the EID of Bob's new cell phone,
all UIA devices now treat this \emph{link} record
as logically referring to all of Bob's merged root namespaces
as defined according to his \emph{merge} records,
giving Alice a convenient name for Bob's implicit ``user identity.''
Any UIA device now treats records
affecting any of Bob's merged root namespaces
as affecting all of them,
making Alice's \emph{link} record for {\tt Bob} in effect
name Bob {\em as a user} rather than Bob's cell phone.
To resolve the name {\tt laptop.Bob} on Alice's iPod, for example,
the iPod follows Alice's \emph{link}
to the root namespace of Bob's cell phone,
then follows Bob's \emph{name device} link for {\tt laptop}
to find the laptop's EID,
even though the latter \emph{name device} record was originally written
with reference to Bob's laptop's root namespace.

By identifying users
indirectly via their clusters in this way,
UIA avoids imposing on users
the burden of having to manage any kind of explicit {\em user identifiers},
or the per-user cryptographic key pairs that would presumably be needed
to generate such identifiers securely.
UIA can therefore give logical identities and meaningful names
to both users and devices,
but only devices actually need to have explicit identifiers,
which they can create automatically for themselves.

\subsection{Groups}

Though not illustrated in the figure, UIA easily supports shared
groups as well.  For example, the members of a household may list 
their common devices under the suffix {\tt.home} by creating a 
new namespace on each device, linking from the device's root namespace
to the new namespace via the name {\tt home}, and then merging all of
the new namespaces together.  Moreover,
the devices need not all merge directly to each other;
a single spanning tree suffices
to join all the devices' respective {\tt home} namespaces.

\subsection{Resolving Conflicts}

By merging his home and cell phones at Time~4, Bob creates a
conflict in his cluster, as the name {\tt phone} is now mapped to two
different EIDs, and he renames one device {\tt cell} to resolve it.
To accomplish this renaming, UIA logs a new \emph{name device} record
for the name {\tt cell}, followed by a \emph{remove name} record
pointing to the original \emph{name device} record for that EID.
The \emph{remove name} record
does not actually delete the original \emph{name device} record,
but merely causes all devices that see it
to ignore the original \emph{name device} record
for purposes of name resolution and conflict detection.
Since there is now only one ``active'' \emph{name device} record
for the name {\tt phone} in Bob's root namespace,
the conflict is effectively resolved on each device
as soon as that device obtains the \emph{remove name} record via gossip.

\com{	Just try to clarify the above a bit more instead... -baf
\footnote{Note that the use of \emph{remove name}
record, (as well as the \emph{unlink} and \emph{stop merge} records)
requires slight modifications to the name resolution procedure describe
above, although we omit the details for brevity.} It does not matter
from which of his devices Bob issues the rename command, since all his
devices interpret the new records the same way after obtaining them via
gossip.
}

\subsection{Lost or Stolen Devices}

To remove the lost cell phone from Bob's cluster at Time~6 in the
figure, his laptop logs a \emph{stop merge} record pointing to the last
known legitimate entry in the cell phone's log.  This record instructs
Bob's devices to continue using old names that the cell phone created,
while ignoring any subsequent records that may be created by the phone,
if the cell phone falls into the hands of a thief for example.
To prevent
the cell phone from introducing conflicting versions of old records,
each UIA log record contains the secure hash of the record that
preceded it.  The secure hash of a single record thus secures an entire
prefix of the log. 

Bob's laptop also logs a {\em remove name} record to delete the cell
phone's EID from his namespace.  The pointer to the initial creation
record in the remove record allows Bob to reuse the name ``cell'' later
with no ambiguity as to which one of the {\em name device} records
the {\em remove name} record actually refers to.

\com{	-- shortened and moved to section 3, though still not sure... -baf
If Bob does not promptly remove the stolen phone from his
cluster, the thief might unfortunately gain access
to Bob's other devices through it, an inherent risk of
greater connectivity.  Standard login mechanisms or biometric security
features, increasingly common on smart portable devices, can help
mitigate this risk, and UIA's append-only logs help facilitate recovery
if the intruder modifies UIA state.
}

\subsection{Routing}

In order to gossip, devices must be able to communicate.  To this end,
they keep track of the IP addresses of others to which they have been
introduced in a local table.  To find a peer that is no longer reachable
at its last known IP address,
a device uses a Gnutella-like expanding ring search
through the network of its reachable peers.
Each device also remembers old IP addresses
to which a peer might have returned in case a search is
unsuccessful.

A successful search result includes the full path taken by the search.
If a device cannot establish a direct IP-level connection
to a peer itself---perhaps
because the peer is behind a NAT or firewall and cannot accept incoming
connections---the device asks the next-to-last node in the search path
to forward traffic to the peer on its behalf.  The devices then use this
channel to establish a direct connection by ``punching a hole''
through the NAT~\cite{ford05p2p} if possible;
otherwise they continue communicating through the intermediary.

Since UIA devices can route opportunistically
through their social neighbors,
the ability of any two frequently-moving devices on the Internet
to locate each other reliably
does not depend on any centralized or manually-configured servers,
but only on the existence of {\em some} UIA device
somewhere in the two devices' common ``social neighborhood''
with an accessible and relatively stable IP address.
This ``rendezvous device'' could happen to be a public server of some kind,
perhaps even one specifically set up to help other UIA devices rendezvous,
but UIA does not depend on this being the case;
the rendezvous device could just as well be a friend's home PC
attached via DSL a ``well-behaved'' or suitably configured NAT.

Each UIA device by default actively monitors the IP addresses
only of its {\em immediate} social neighbors---%
i.e., the user's own devices and those of his ``first-degree'' contacts.
Since most users are expected to have tens
or at most a few hundred immediate contacts,
each node's routing traffic burden should be manageable,
and thus the system should scale well
even if the total size of the interconnected UIA network
is orders of magnitude larger.
The time required to locate an arbitrary device
naturally increases with ``social distance,''
and its chance of success similarly decreases,
but this property is appropriate since we expect people to use UIA
mostly to communicate within their immediate social neighborhood.
Nonetheless,
efficient routing to arbitrary EIDs is an important direction for future
work.

\section{Implementation}
\label{sec-impl}

A protoype UIA implementation currently runs on Linux and Mac OS~X.
This prototype is divided into separate routing and naming layers, both
of which run as user-level daemons
to which UIA-aware applications on the device
can directly interface via Sun RPC.
Through these interfaces,
a UIA-aware application can send packets to EIDs of its
choice, listen for packets on its own EID, discover potential peers on
the local-area network, enter new names into its namespace, and resolve
UIA names to EIDs.
The prototype uses 
Apple's Bonjour library for local-area device discovery and
SSL for secure communication between peers.

Furthermore, the UIA prototype provides support for legacy applications
that support IPv6.  We have successfully used Apache, Firefox, and
OpenSSH over UIA, without modification or even recompilation, via this legacy
interface.  For routing, UIA uses the \emph{tun} device to disguise EIDs
as IPv6 addresses.  In this way, applications can bind a socket to the
local device's EID or connect to a remote device by EID.  For naming,
UIA provides a local DNS proxy that sends each lookup request to both
the UIA naming layer and the device's normal DNS server and returns a
combined result.

Currently, users perform device discovery and introduction using
command-line programs.  A graphical user interface is under development.

%  \note{Do we want to say anything
%else about that?  Maybe an example of some functionality not available
%through the legacy interface?}

%\subsection{Interaction with Existing Infrastructure}

%\note{UIA names look just like DNS names.  As a result,
%we can hack \emph{gethostbyname} to look up each name in both systems
%and return a combined result.  We can also prevent the creation of
%namespaces like ``.com'', ``.org'', etc.~to prevent obvious collisions.}

%\note{Can gossip log records through a DHT, store them in some
%centralized server, etc.}

%\note{Can use a centralized server or DHT to help with routing, etc.}

%\note{Cite: HIP, i3, UIP, ??.}

\section{Related Work}
\label{sec-related}

Existing Internet mobility mechanisms require configuration effort
and technical expertise that deters even many sophisticated users.
Dynamic DNS~\cite{rfc2136} supports automatic IP address updates,
but devices still become inaccessible when
behind a NAT~\cite{rfc3027}.  Mobile IP~\cite{rfc3344} gives a mobile
device the illusion of a fixed IP address, but requires a dedicated
forwarding server at a static, public IP address.
UIA in contrast relies on user-relative names, self-configuring EIDs,
and opportunistic use of peer devices
for rendezvous and traffic forwarding to support mobility.

Ad hoc naming schemes such as Bonjour~\cite{bonjour} allow
devices to choose their own names on local-area networks, but
these names are insecure and ephemeral: any device joining a network can
claim any unused name, and a device's name becomes invalid
as soon as it moves to a different network.  UIA uses the Bonjour
libraries to discover new devices on the local network, 
but UIA
names persist and remain securely bound to the original target device
despite later migration of the devices involved.

UIA's user-relative naming model may be useful to other
systems that use cryptographic host identifiers,
such as HIP~\cite{moskowitz03hip-arch}, $i3$~\cite{stoica02internet},
and SFR~\cite{balakrishnan2003semantic-free}.  
Though UIA takes advantage of global infrastructure when available,
UIA does not depend on it
and can continue providing naming and communication among local devices
even when disconnected from the Internet.

% This next sentence is not true according to my understanding of i3.  
% --srhea
%
%Since any $i3$ participant can listen on any name of its choice,
%$i3$ does not provide users an easy way
%to authenticate their mobile devices to each other,

UIA's user-relative naming model is inspired in part by
SDSI/SPKI~\cite{rivest96sdsi,rfc2692,rfc2693}.  Like SDSI, UIA identifies devices by their
public keys, and allows users to define relative names.  
SDSI's model for designated certificate servers does not adapt well
to disconnected mobile devices, however.
UIA also simplifies key management
by identifying users implicitly via their personal clusters
instead of requiring them
to manage per-user public/private key pairs explicitly,
and UIA handles lost or stolen devices without
rekeying and thus losing the user's identity.

UIA's relaxed consistency model is partially inspired by
Bayou~\cite{terry95managing} and Ivy~\cite{muthitacharoen02ivy}.
The semantics of UIA's log records do not require devices
to converge on a single total ordering of their logs, however,
simplifying conflict detection and resolution.

Finally, UIA is a continuation of work begun with the Unmanaged Internet
Protocol~\cite{ford03uip, ford03scalable}.  Unlike the earlier work, however,
UIA routes along overlay links that mirror the social trust
relationships of the devices involved, as in Turtle~\cite{turtle} and
SPROUT~\cite{sprout}.

\section{Conclusion}
\label{sec-concl}

UIA is a connectivity architecture
that facilitates global peer-to-peer sharing
of information and services
between personal devices
by giving technically unsophisticated users
a simple, intuitive, and secure way to name
both their devices and other users.
By combining
local device introduction and user-relative naming
with self-certifying global device identities
and globally distributed personal namespaces,
UIA represents a unique application of social networking concepts
to the problem of ad hoc peer-to-peer naming.

\com{	-- too redundant with related work, as Frans points out. -baf
Beyond its intended usage model,
UIA's naming scheme may also be applicable
to providing easy-to-use human-readable names
in other peer-to-peer connectivity architectures,
such as HIP~\cite{moskowitz03hip-arch}, $i3$~\cite{stoica02internet},
and SFR~\cite{balakrishnan2003semantic-free}.
}

\com{

Users normally assign UIA names by introducing devices to each other
on a common local-area network,
but the resulting names persist and remain usable
as the devices migrate elsewhere.
All of a user's devices share a common distributed namespace,
and different users can share access to their namespaces
and refer to each other's devices via convenient user-relative names
analogous to ``Alice's desktop.''
Most importantly,
UIA naming does not require users
to have technical knowledge of protocols or IP addresses
or to depend on central administrative authorities
for setup or maintenance.

} %com

\section*{ACKNOWLEDGEMENTS}

This research is sponsored by the T-Party Project, a joint research
program between MIT and Quanta Computer Inc., Taiwan, and by the
National Science Foundation under Cooperative Agreement No. ANI-0225660
(Project IRIS). 

%\begin{scriptsize}
\bibliographystyle{plain}
\bibliography{uia}

%\end{scriptsize}

\end{document}